\documentclass{webofc}
\usepackage[varg]{txfonts} 
\usepackage{lineno}

\begin{document}
\title{Elliptic flow of strange and multi-strange hadrons in isobar collisions at $\sqrt{s_{\mathrm {NN}}}$ = 200~GeV at RHIC}

\author{\firstname{Priyanshi} \lastname{Sinha}\inst{1}\fnsep\thanks{\email{priyanshisinha@students.iisertirupati.ac.in} }(for the STAR Collaboration)
}
             

\institute{Indian Institute of Science Education and Research (IISER) Tirupati, India}

\abstract{%
  We report the elliptic flow ($v_{2}$) of $K_{s}^{0}$, $\Lambda$, $\bar{\Lambda}$, $\phi$, $\Xi^{-}$, and $\overline{\Xi}^{+}$ at mid-rapidity for Ru+Ru and Zr+Zr collisions at $\sqrt{s_{\mathrm {NN}}}$ = 200 GeV. The transverse momentum ($p_{T}$) and centrality dependence of $v_{2}$ have been studied. The number of constituent quark (NCQ) scaling for all the (multi-)strange hadrons has also been tested. The dependence in the ratio of average $v_{2}$ ($\langle$$v_{2}$$\rangle$) on centrality shows a deviation from unity indicating a difference in nuclear structure and deformation between the isobars. The results show a systematic size dependence when compared to Cu+Cu, Au+Au, and U+U collisions at similar beam energies.
}
\maketitle
\section{Introduction}
\label{intro}
A dedicated isobar collision run of $^{96}_{44}$Ru+$^{96}_{44}$Ru and $^{96}_{40}$Zr+$^{96}_{40}$Zr at $\sqrt{s_{\mathrm {NN}}}$ = 200 GeV was successfully carried out at RHIC in 2018 to measure the charge separation along the magnetic field, called the Chiral Magnetic Effect (CME)~\cite{isobarData}. These collisions are considered to be an effective way to minimize the flow-driven background contribution in search for the possibly small CME signal~\cite{expProp, ReviewCME}. The recent studies also show probing of nuclear structures via $v_{2}$ ratios as well as the $v_{2}$-[$p_{T}$] correlations in isobar collisions~\cite{betaDeform, v2ptcorrelation}. The deformation parameters are different between the two species and flow measurements are highly sensitive to them. Strange and multi-strange hadrons have a small hadronic cross-section compared to light hadrons, making their elliptic flow an excellent probe for understanding the initial state anisotropies of these isobar collisions.

\section{Analysis details}
\label{sec-1}
In these proceedings, we report strange and multi-strange hadron $v_{2}$ in Ru+Ru and Zr+Zr collisions at $\sqrt{s_{\mathrm {NN}}}$ = 200 GeV using the data collected by the STAR experiment.  A total of nearly 650M events have been analysed for both the isobar collisions. $\phi$-mesons have been reconstructed using the invariant mass technique through its hadronic decay channel: $\phi \rightarrow K^{+} K^{-}$ ($48.9 \pm 0.5\%$) in midrapidity $\left|y\right|$ $<$ 0.5~\cite{pdg}. Event mixing technique is used for combinatorial background estimation. 
The weakly decaying neutral strange particles $K_{s}^{0}$ and $\Lambda(\bar{\Lambda})$ are reconstructed using invariant mass technique and their weak-decay (V0) topology through the decay channel: $K_{s}^{0} \rightarrow \pi^{+}  + \pi^{-}$ ($69.20 \pm 0.05\%$) and $\Lambda \rightarrow p  + \pi^{-}$ ($63.9 \pm 0.5\%$), respectively~\cite{pdg, Strange}. The multi-strange particle $\Xi^{-}(\overline{\Xi}^{+})$ decays into a charged daughter and a neutral V0 particle ($\Lambda$), which in turn decays into two charged particles. Their reconstruction involves finding two secondary vertices and the various topological selections. The combinatorial background for the weakly decaying particles is constructed using rotational background method~\cite{multiStrange}. The $\eta$-sub event plane method with an $\eta$ gap of 0.1 has been used to calculate $v_{2}$ of these (multi-)strange hadrons~\cite{Strange}. The maximum EP resolution of nearly 48\% is achieved for both the collision systems.  

\section{Results}
\label{sec-2}
\begin{figure}[h!]
\centering
\begin{tabular}{cc}
 \includegraphics[width=5.5cm]{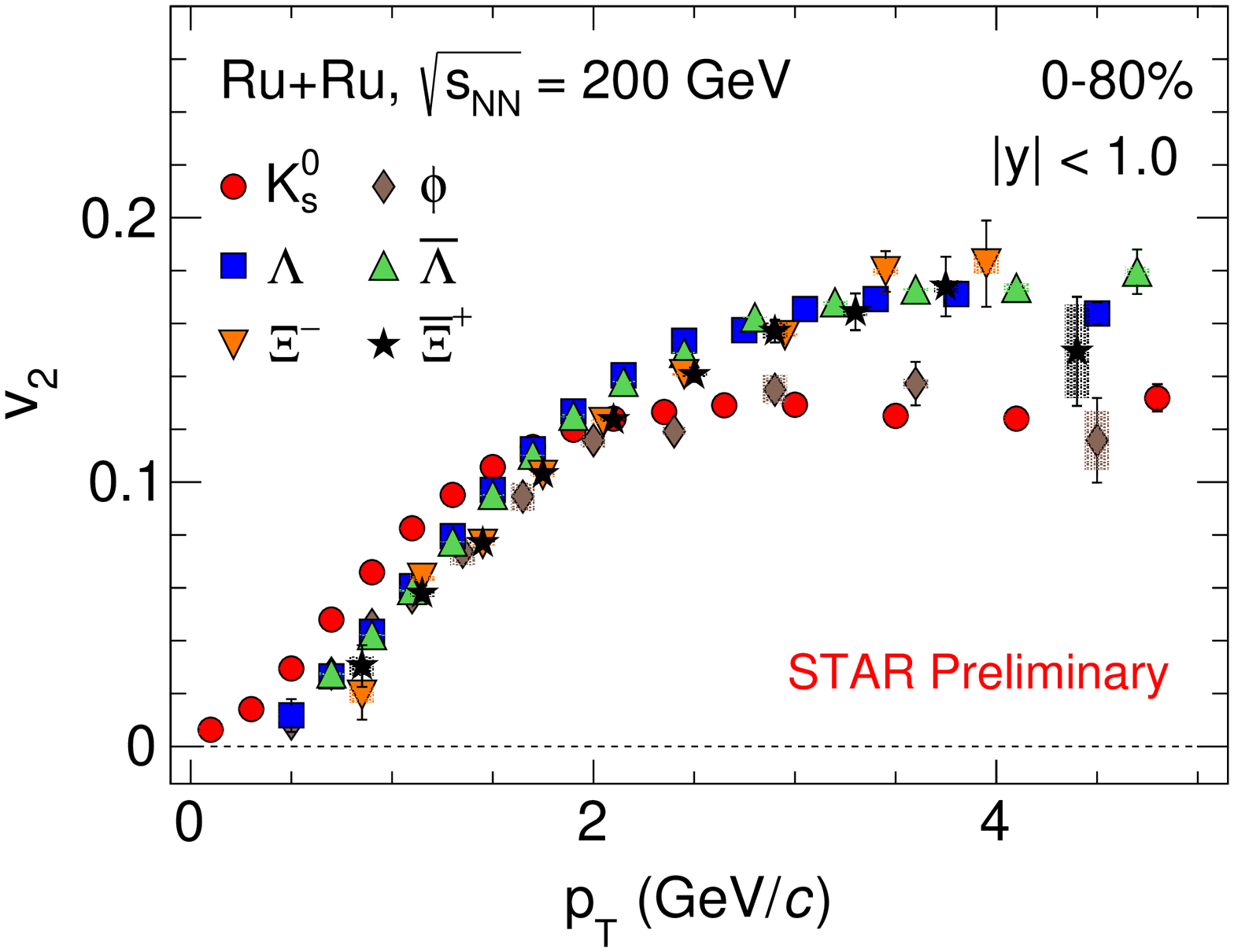} &
 \includegraphics[width=5.5cm]{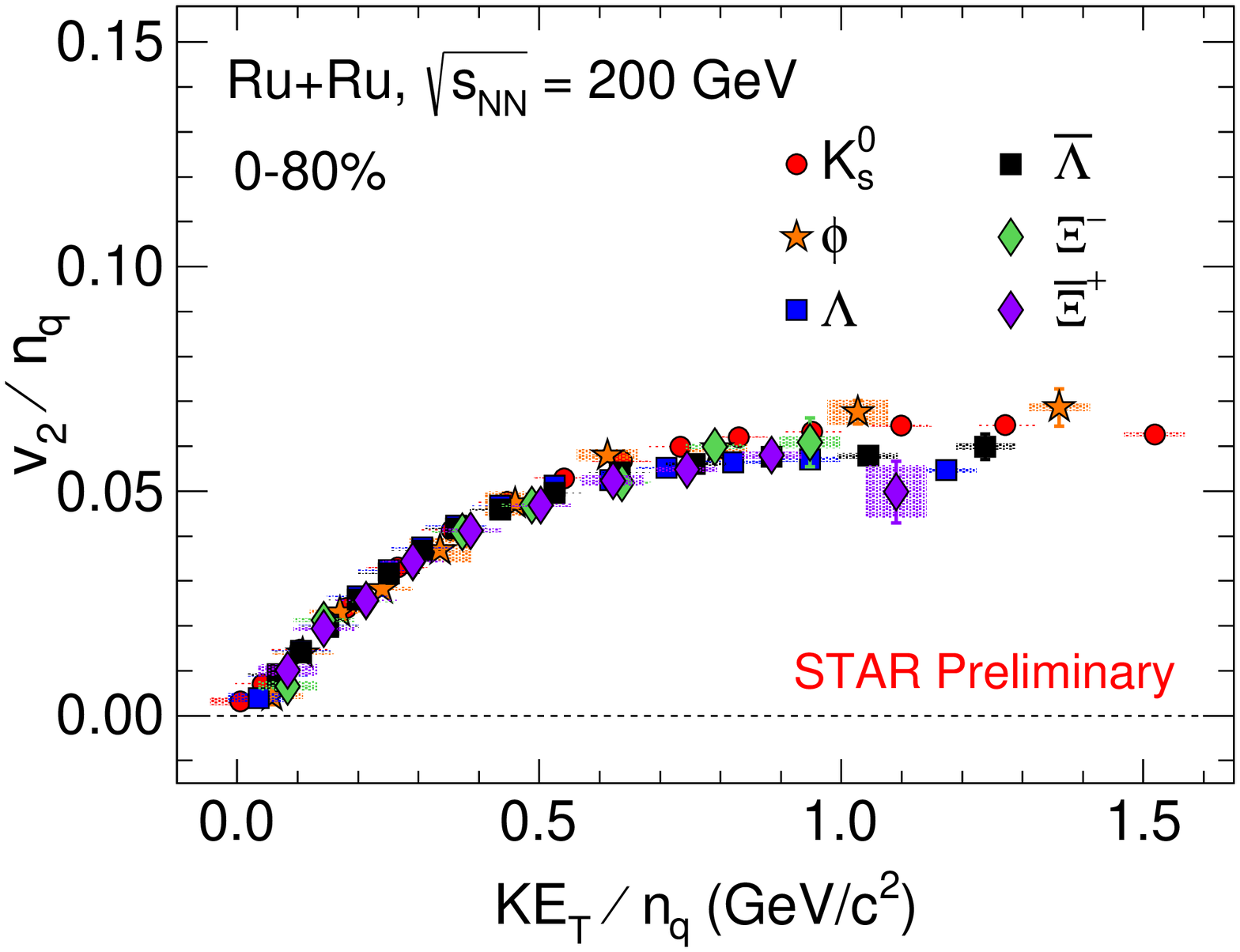} 
\end{tabular}
\vspace{-0.2cm}
\caption{Left panel: $v_{2}$ as a function of $p_{T}$ of (multi-)strange hadrons; Right panel: NCQ-scaled $v_{2}$ as a function of transverse kinetic energy for Ru+Ru collisions at $\sqrt{s_{\mathrm {NN}}}$ = 200 GeV. The vertical lines and shaded boxes denote statistical and systematic uncertainties, respectively.}
\vspace{-0.3cm}
\label{fig:Fig.1}
\end{figure}

\begin{figure}[h!]
\centering
\begin{tabular}{cc}
 \includegraphics[width=5.5cm]{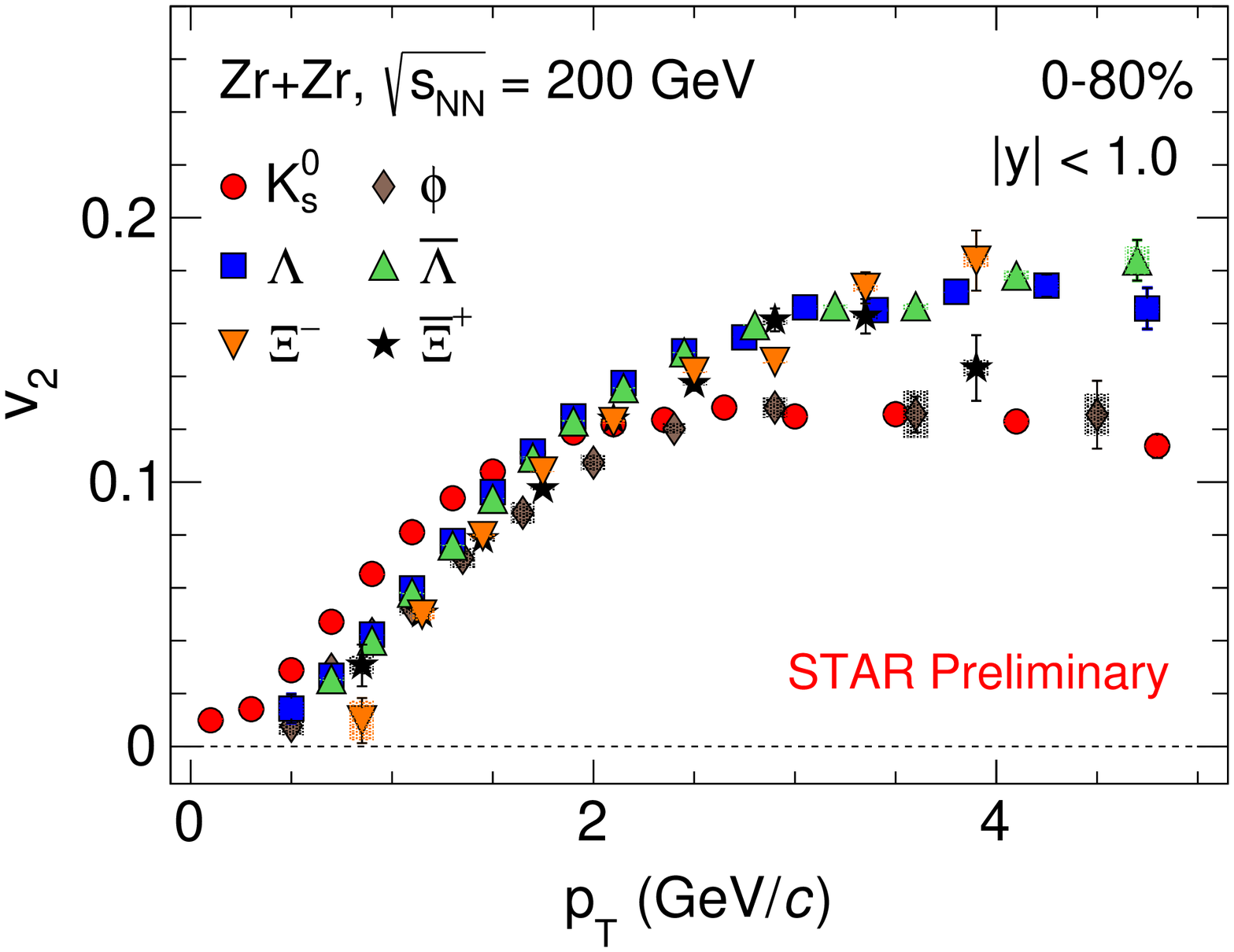} &
 \includegraphics[width=5.5cm]{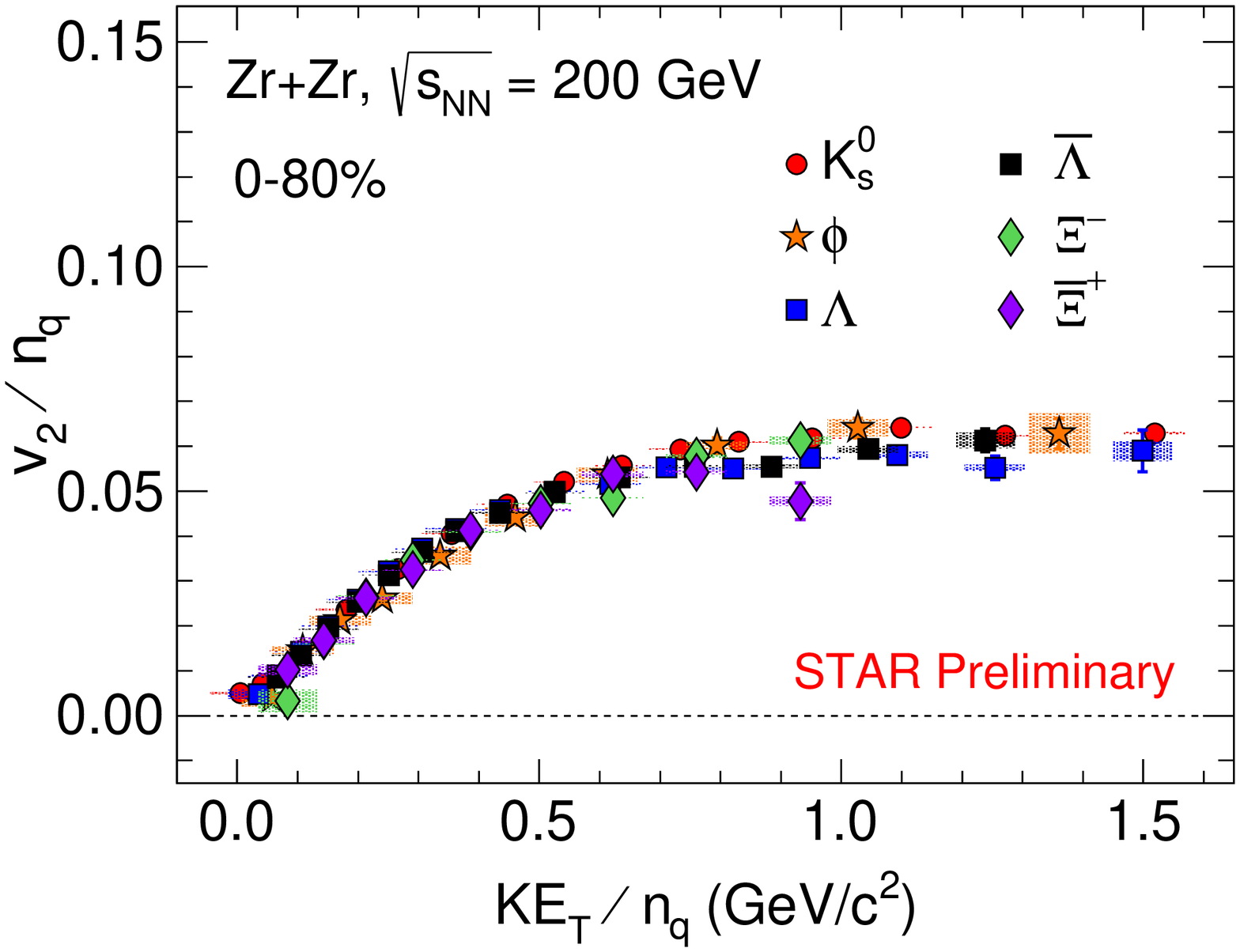} 
\end{tabular}
\vspace{-0.2cm}
\caption{Left panel: $v_{2}$ as a function of $p_{T}$ of strange hadrons; Right panel: NCQ-scaled $v_{2}$ as a function of transverse kinetic energy for Zr+Zr collisions at $\sqrt{s_{\mathrm {NN}}}$ = 200 GeV. The vertical lines and shaded boxes denote statistical and systematic uncertainties, respectively.}
\vspace{-0.2cm}
\label{fig:Fig.2}
\end{figure}

\begin{figure}[h!]
\centering
\begin{tabular}{cc}
\includegraphics[width=5.5cm]{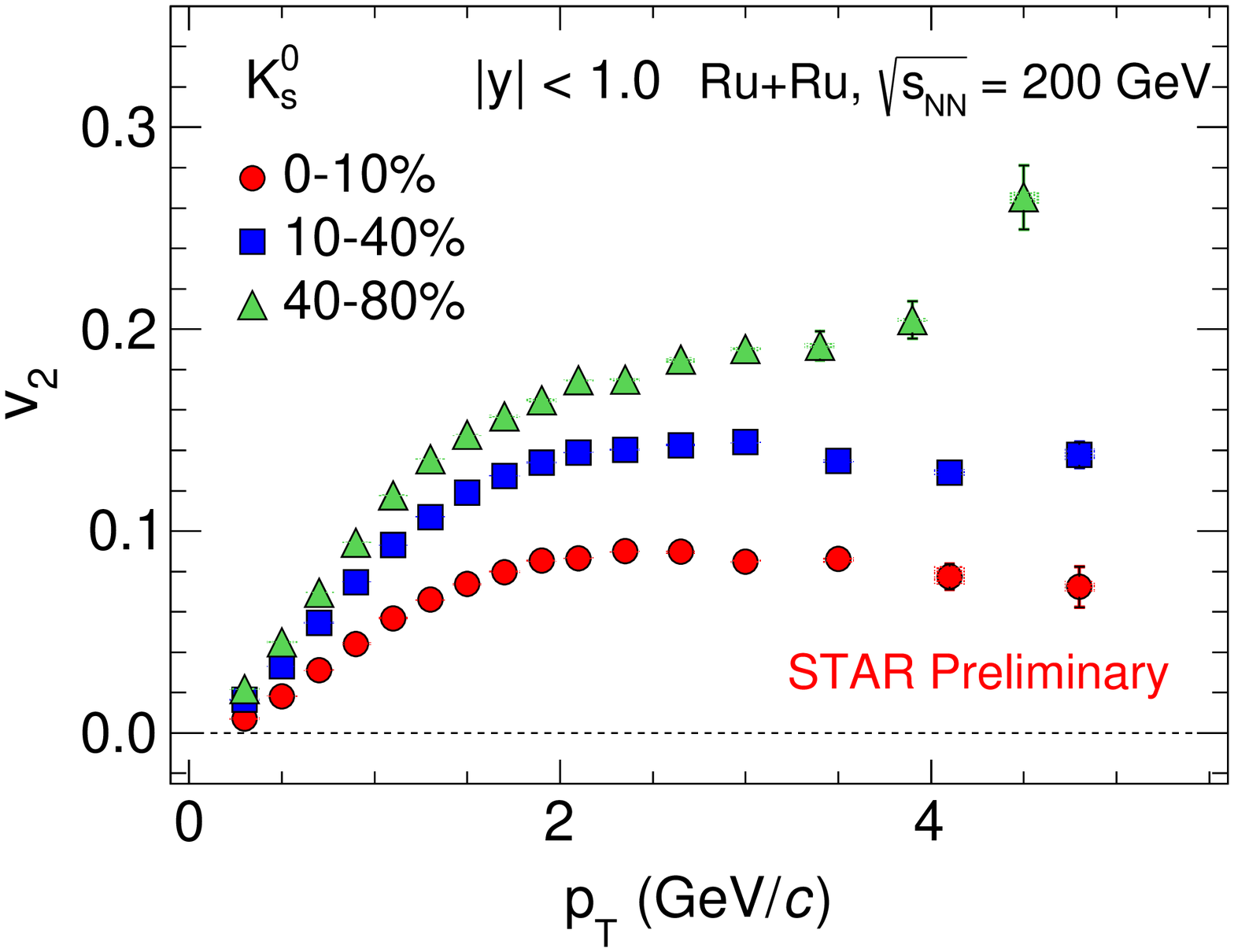} &
 \includegraphics[width=5.5cm]{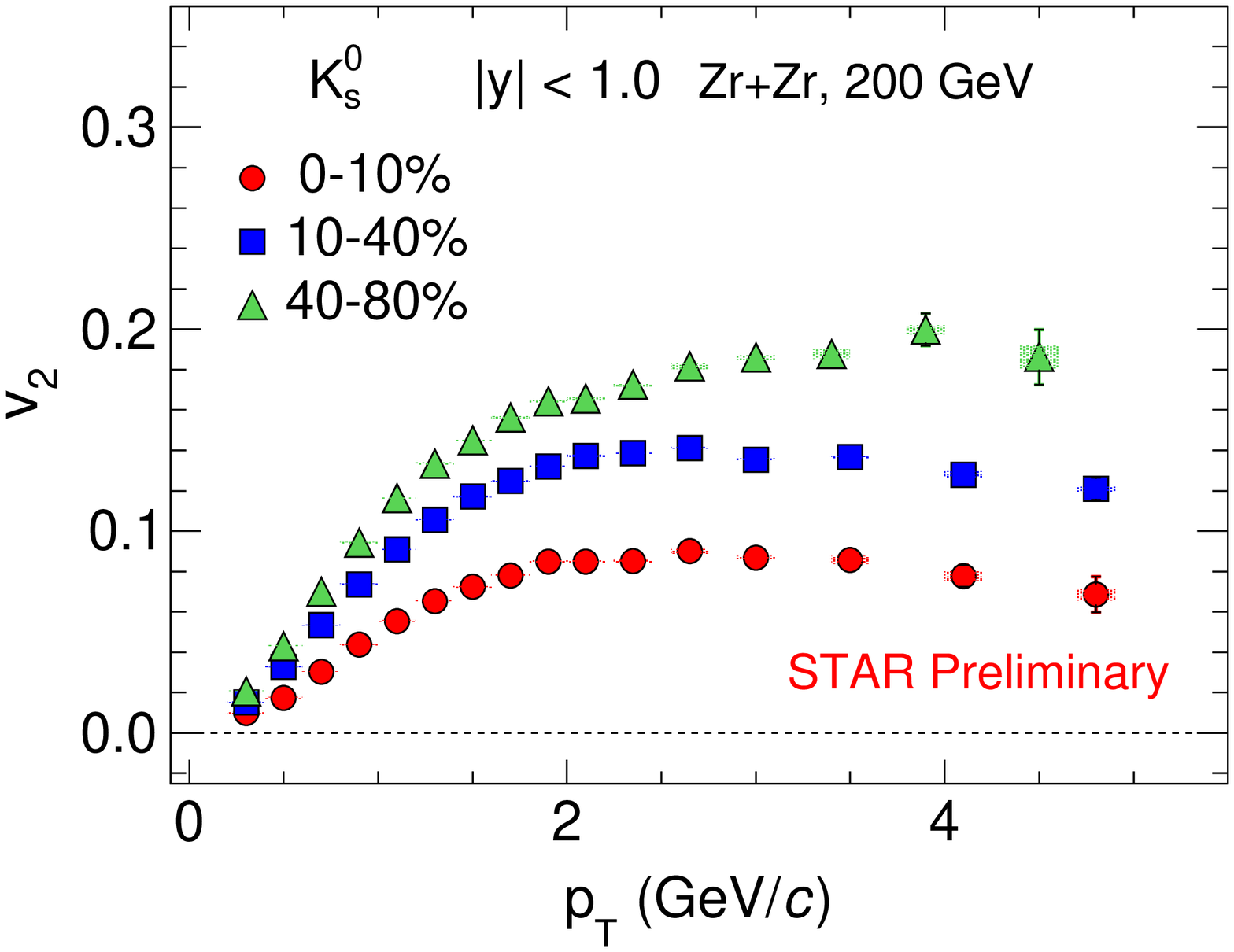} \\
 \includegraphics[width=5.5cm]{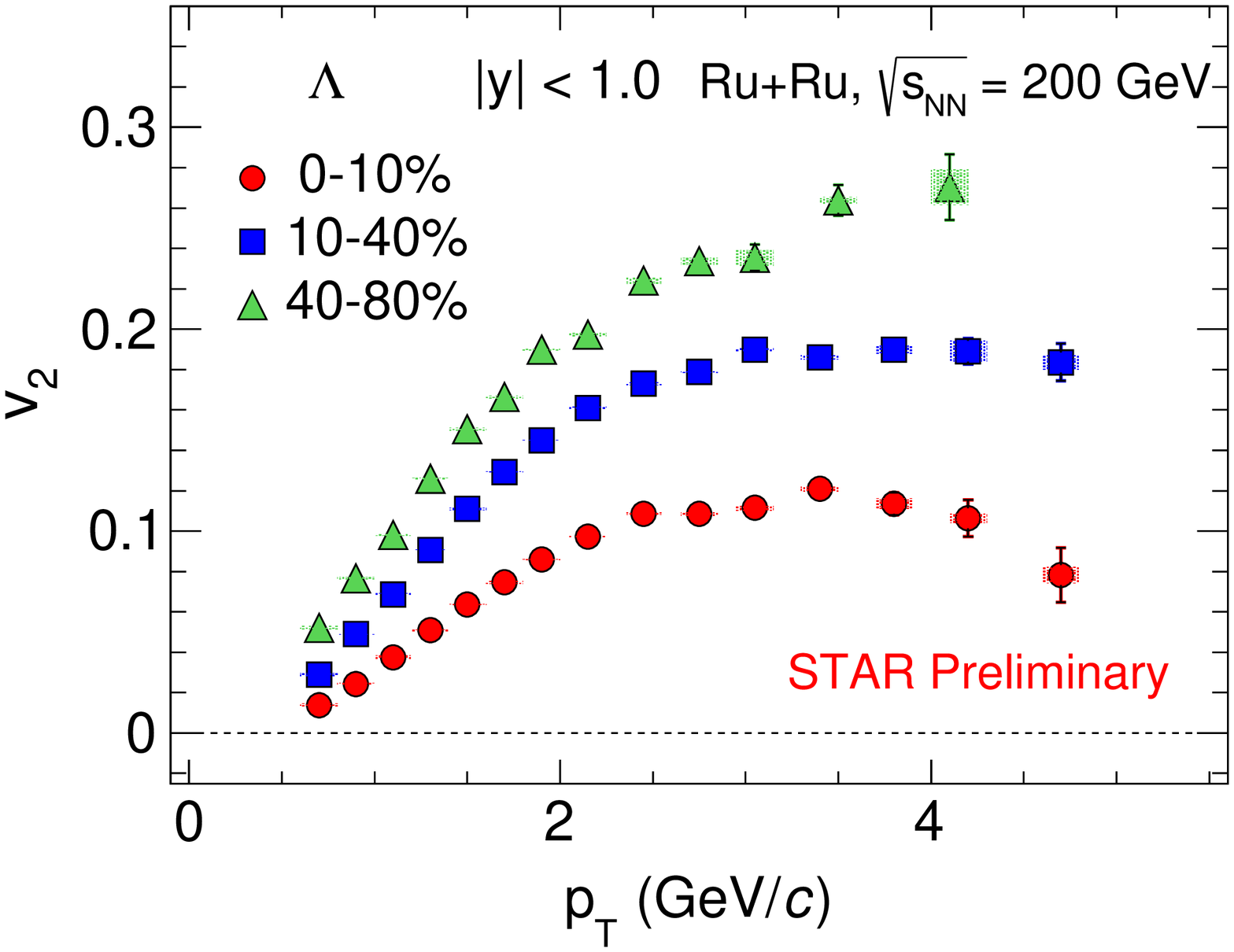} &
 \includegraphics[width=5.5cm]{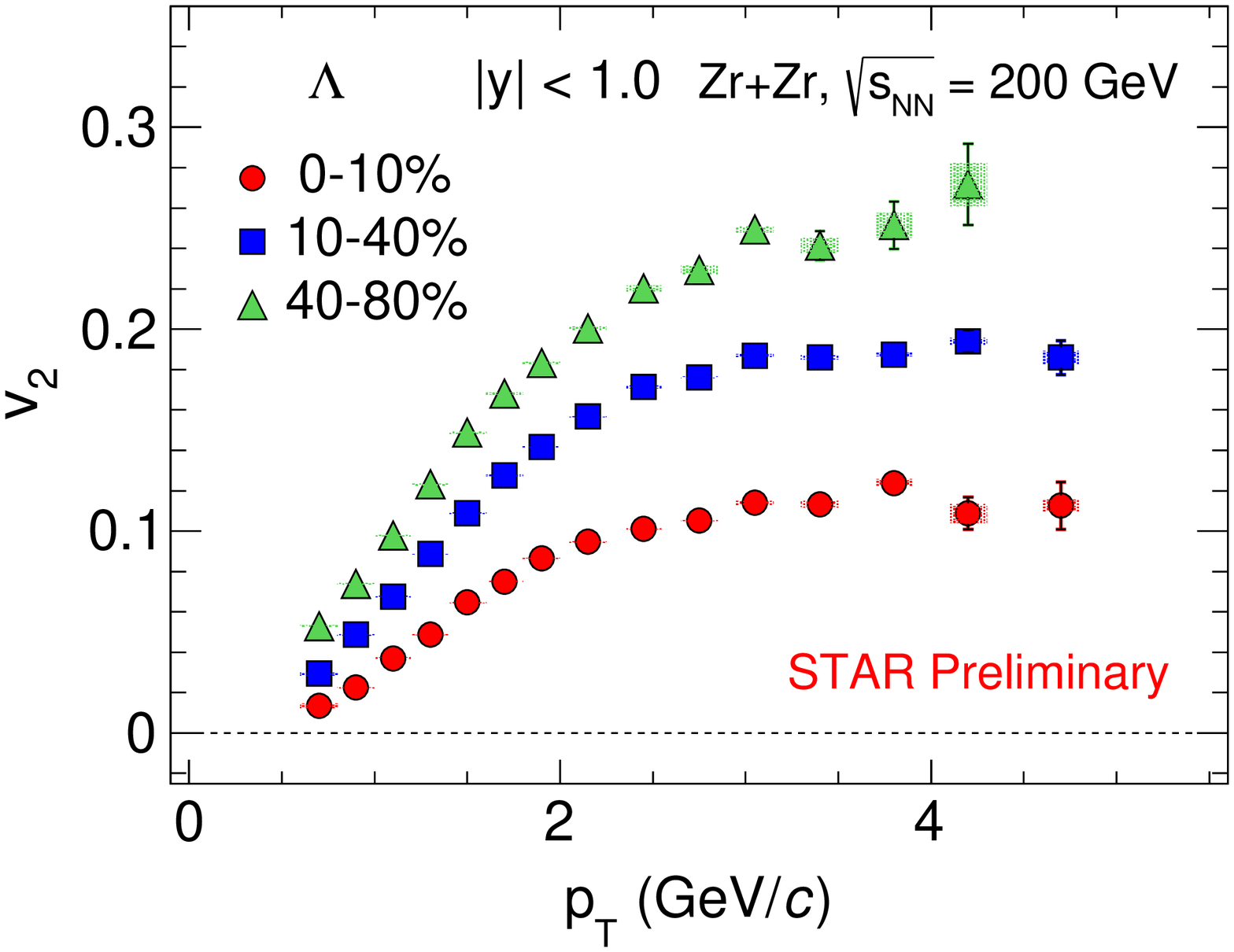} \\
  \includegraphics[width=5.5cm]{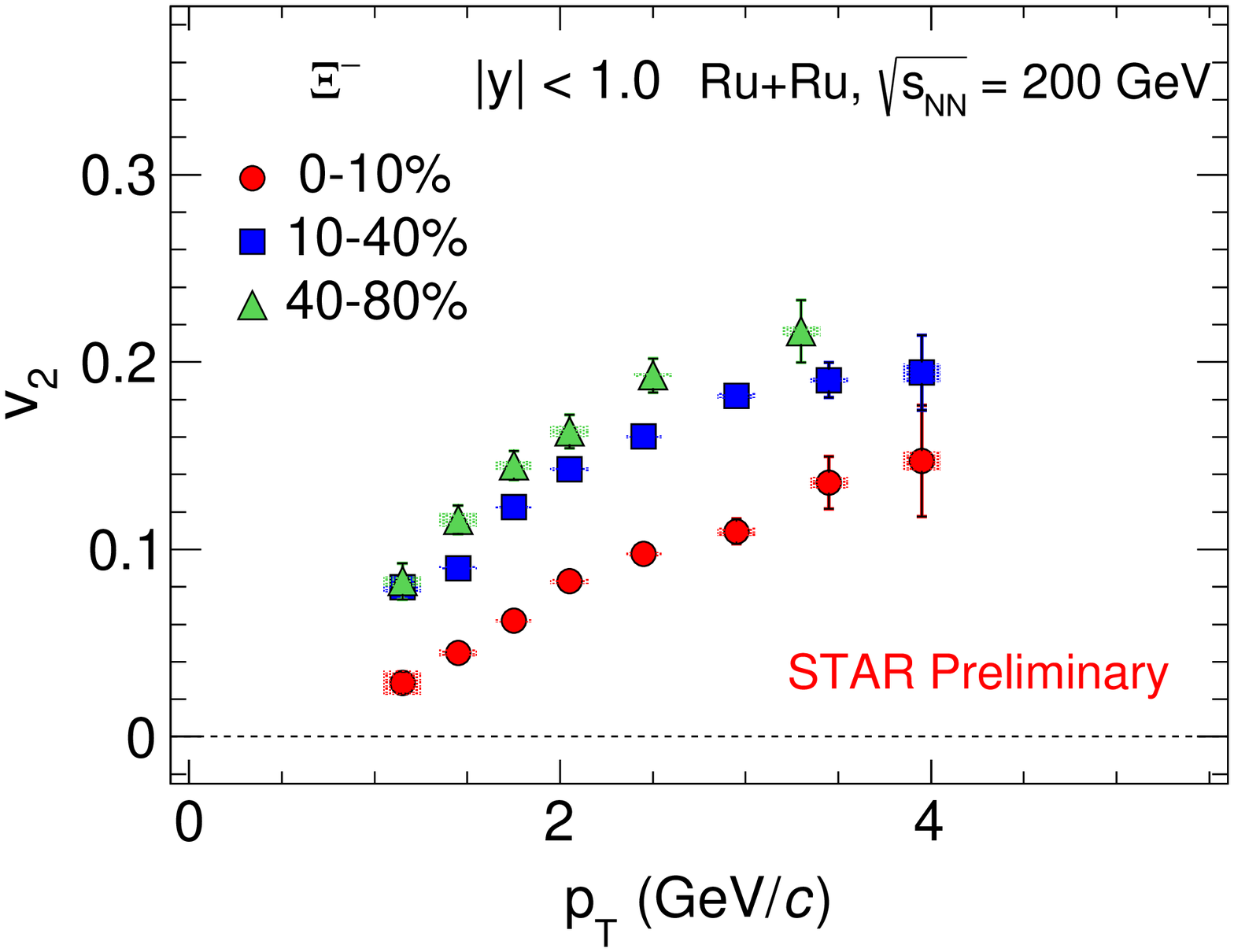} &
 \includegraphics[width=5.5cm]{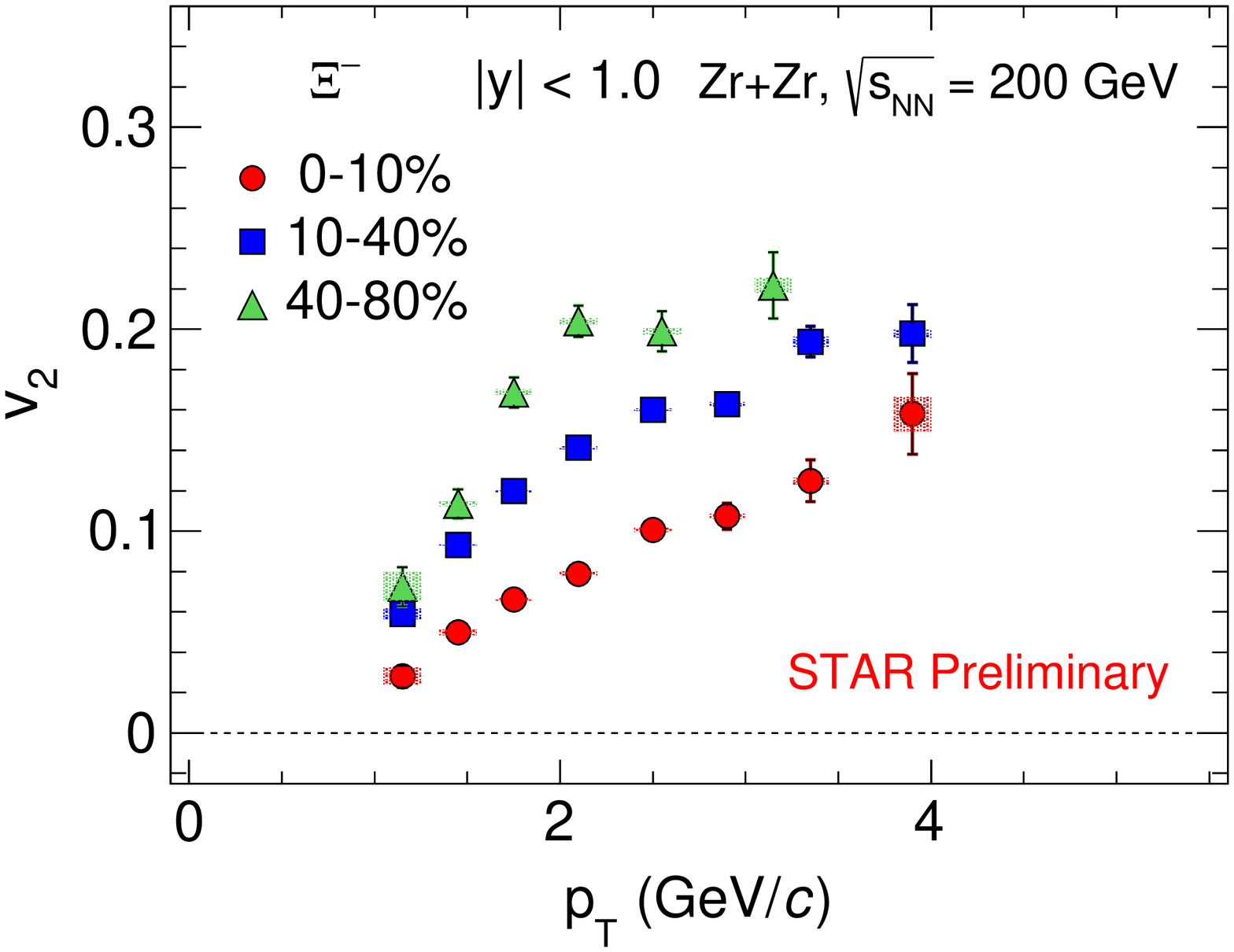} 
\end{tabular}
\vspace{-0.2cm}
\caption{Left panel: Centrality dependence of $v_{2}$ of $\Lambda$ as a function of $p_{T}$ in Ru+Ru collisions; Right Panel: Same for Zr+Zr collisions at $\sqrt{s_{\mathrm {NN}}}$ = 200 GeV. The vertical lines and shaded boxes denote statistical and systematic uncertainties, respectively.}
\vspace{-0.5cm}
\label{fig:Fig.3}
\end{figure}

\begin{figure}[h!]
\centering
 \includegraphics[width=12cm]{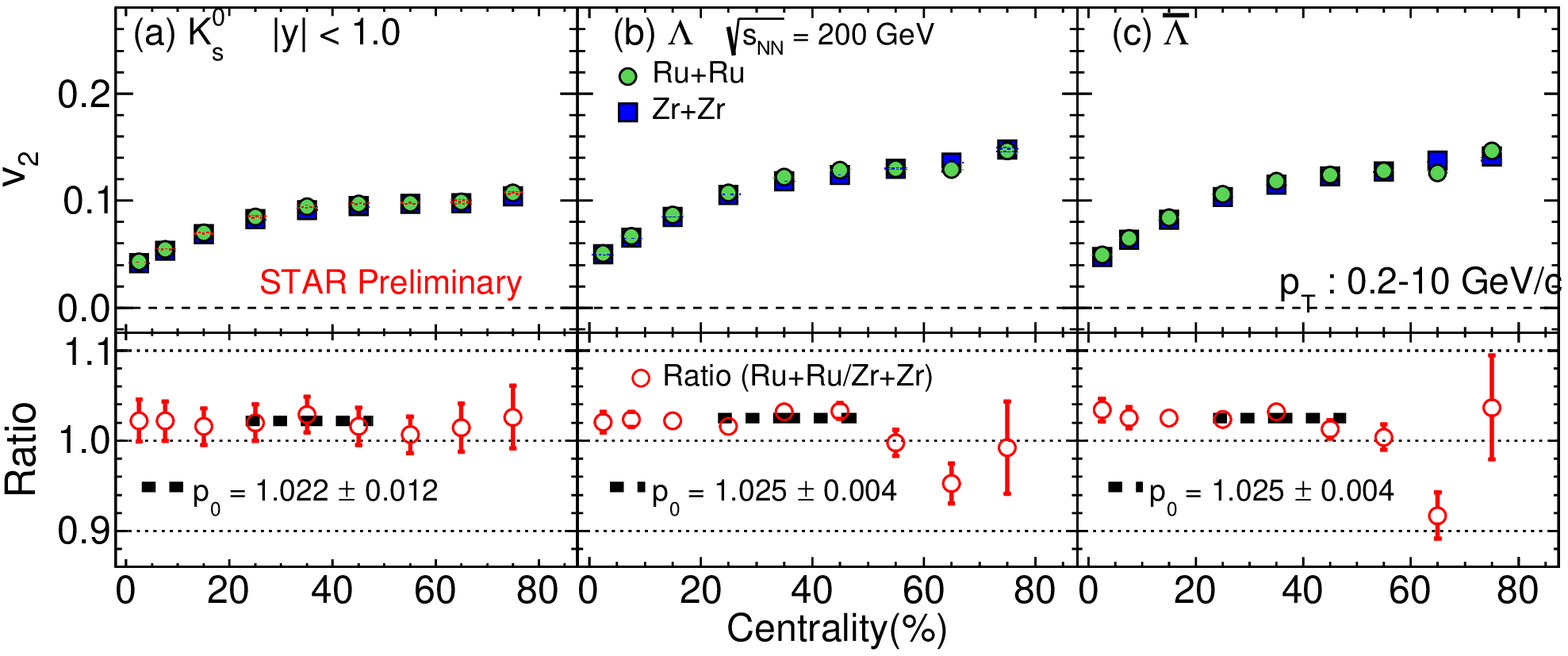} 
 \vspace{-0.25cm}
\caption{$p_{T}$-integrated $v_{2}$ as a function of centrality for $K_{s}^{0}$, $\Lambda$, and $\bar{\Lambda}$ in Ru+Ru and Zr+Zr collisions at $\sqrt{s_{\mathrm {NN}}}$ = 200 GeV. The vertical lines on the ratio includes statistical and systematic uncertainties. The dotted lines denotes the fitting with a constant.}
\label{fig:Fig.4}
\end{figure}

The left panels of Figs.~\ref{fig:Fig.1} and~\ref{fig:Fig.2} show the $v_{2}$ of strange and multi-strange hadrons as a function of $p_{T}$ for minimum bias Ru+Ru and Zr+Zr collisions at $\sqrt{s_{\mathrm {NN}}}$ = 200 GeV, respectively. An approximate mass ordering at low $p_{T}$ and a baryon-meson splitting at intermediate $p_{T}$ have been observed. All particles and anti-particles tend to follow the number of constituent quark (NCQ) scaling within 10\% shown in the right panel of Fig.~\ref{fig:Fig.1} and~\ref{fig:Fig.2}, indicating the partonic collectivity as well as domination of quark coalescence mechanism during hadronization at intermediate $p_{T}$-region. 
A clear centrality dependence of $v_{2}$ has been observed for $K_{s}^{0}$, $\Lambda$ and $\Xi^{-}$ as shown in Fig.~\ref{fig:Fig.3} for the isobar collision systems. 
We also study the $p_{T}$-integrated $v_{2}$ for strange hadrons as a function of the collision centrality in the isobar collisions as shown in Fig.~\ref{fig:Fig.4}. The ratios of $v_{2}$ between the two isobar collisions for $K_{s}^{0}$, $\Lambda$, and $\bar{\Lambda}$ are comparable with charged hadron data and show a deviation of nearly 2\% from unity in mid-central collisions, indicating a difference in nuclear structure and shape~\cite{isobarData}.
\begin{figure}[h!]
\centering
 \includegraphics[width=13cm]{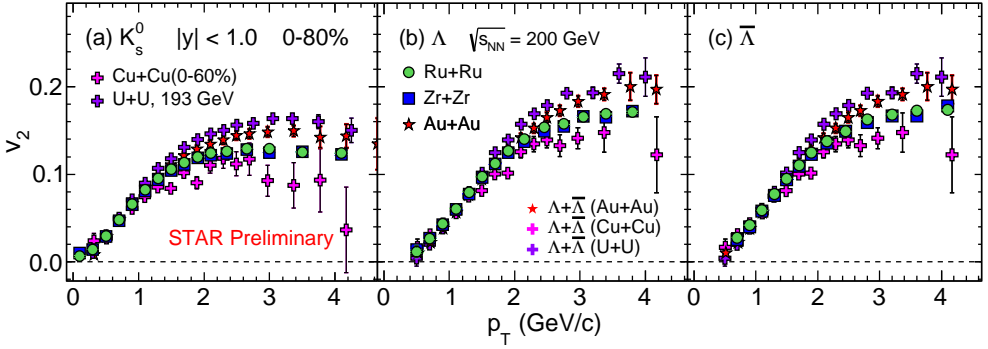} 
 \vspace{-0.75cm}
\caption{$v_{2}$ of strange hadrons in minimum bias Cu+Cu, Ru+Ru, Zr+Zr, Au+Au collisions at $\sqrt{s_{\mathrm {NN}}}$ = 200 GeV and U+U collisions at $\sqrt{s_{\mathrm {NN}}}$ = 193 GeV~\cite{CuData, AuData, UData}.}
\vspace{-0.5cm}
\label{fig:Fig.5}
\end{figure}
We studied the system size evolution of $v_{2}$ by comparing the minimum bias $^{63}_{29}$Cu+$^{63}_{29}$Cu, $^{96}_{44}$Ru+$^{96}_{44}$Ru, $^{96}_{40}$Zr+$^{96}_{40}$Zr, $^{197}_{79}$Au+$^{197}_{79}$Au collisions at $\sqrt{s_{\mathrm {NN}}}$ = 200 GeV, and $^{238}_{92}$U+$^{238}_{92}$U collisions at $\sqrt{s_{\mathrm {NN}}}$ = 193 GeV~\cite{CuData, AuData, UData}. Figure \ref{fig:Fig.5} shows an approximate system size dependence of $v_{2}$ for $p_{T}$ $>$ 1.8 GeV/$c$, based on the nuclear size. $v_{2}$ in U+U and Au+Au is observed to be higher, whereas that in Cu+Cu is slightly lower than those in isobar collisions. 

\section{Summary}
\label{sec-3}
In summary, we have presented the elliptic flow of $K_{s}^{0}$, $\Lambda$, $\bar{\Lambda}$, $\phi$, $\Xi^{-}$, and $\overline{\Xi}^{+}$ in Ru+Ru and Zr+Zr collisions at $\sqrt{s_{\mathrm {NN}}}$ = 200 GeV. We observed a mass ordering at low $p_{T}$ and a baryon-meson splitting at intermediate $p_{T}$ in both isobar species. All the strange particles and anti-particles follow the NCQ scaling indicative of the partonic degrees of freedom and coalescence picture for hadronization. The $p_{T}$-integrated $v_{2}$ ratio of strange hadrons shows a deviation of about 2\% from unity. We also observed a system size dependence of $v_{2}$ when comparing to different collision systems at similar beam energy. These measurements provide further constraint on the deformation and nuclear density difference in the two isobar nuclei.

\end{document}